\documentclass[a4paper,11pt]{article}
\pdfoutput=1 

\usepackage{jheppub} 

\usepackage[T1]{fontenc} 

\usepackage{graphicx}
\usepackage{dcolumn}
\usepackage{bm}
\usepackage{slashed}
\usepackage{amsmath,graphicx}
\usepackage[colorlinks=true,linktocpage=true,linkcolor=blue,citecolor=blue]{hyperref}
\usepackage{float}
\usepackage{nicefrac}
\usepackage[normalem]{ulem}
\usepackage{subfigure} 
\usepackage{bbold} 
\usepackage[makeroom]{cancel}

\def\be{\begin{equation}}
	\def\ee{\end{equation}}
\newcommand{\bel}[1]{\begin{eqnarray}\label{#1}}
	\newcommand{\eel}{\end{eqnarray}}
\def\barr{\begin{array}}
	\def\earr{\end{array}}
\def\beq{\begin{eqnarray}}
	\def\eeq{\end{eqnarray}}
\def\bfig{\begin{figure}}
	\def\efig{\end{figure}}

\newcommand{\nn}{\nonumber}
\newcommand{\f}[2]{\frac{#1}{#2}}
\newcommand{\onehalf}{{\nicefrac{1}{2}}}

\newcommand{\p}{\partial}

\newcommand{\tr}{{\rm tr}}
\newcommand{\rf}[1]{Eq.~(\ref{#1})}

\newcommand{\rftwo}[2]{Eqs.~(\ref{#1})~and~(\ref{#2})}
\newcommand{\rfn}[1]{(\ref{#1})}

\newcommand{\rfc}[1]{Ref.~\cite{#1}}

\def\fplusrsxp{f^+_{rs}(x,p)}

\def\fminusrsxp{f^-_{rs}(x,p)}

\def\SmunuU{{\Sigma}^{\mu\nu}}

\def\S0iU{{\Sigma}^{0i}} 
 
\def\SmnU{{\Sigma}^{\mu\nu}}
\def\SmnL{{\Sigma}_{\mu\nu}}
\def\ubarrp{{\bar u}_r(p)}

\def\usp{u_s(p)}

\def\urp{u_r(p)}

\def\vbarrp{{\bar v}_r(p)}

\def\vbarsp{{\bar v}_s(p)}

\def\vsp{v_s(p)}
\def\vrp{v_r(p)}

\def\bmu{\beta_\mu}

\def\umU{u^\mu}

\def\n0{n_{(0)}}
\def\e0{\varepsilon_{(0)}}
\def\P0{P_{(0)}}

\def\omnL{\omega_{\mu\nu}}
\def\omnU{\omega^{\mu\nu}}

\def\oabU{\omega^{\alpha\beta}}
\def\omnLD{{\tilde \omega}_{\mu\nu}}
\def\omnUD{\tilde {\omega}^{\mu\nu}}

\def\epsLmnab{\epsilon_{\mu\nu\alpha\beta}}

\def\epsLmnab{\epsilon_{\mu\nu\alpha\beta}}

\def\pmu{p^\mu}

\def\pv{{\boldsymbol p}}
\def\av{{\boldsymbol a}}
\def\bv{{\boldsymbol b}}
\def\kv{{\boldsymbol k}}

\def\Wxk{{\cal W}(x,k)}
\def\Weqxk{{\cal W}_{\rm eq}(x,k)}
\def\Weqpxk{{\cal W}^{+}_{\rm eq}(x,k)}
\def\Weqmxk{{\cal W}^{-}_{\rm eq}(x,k)}
\def\Weqpmxk{{\cal W}^{\pm}_{\rm eq}(x,k)}

\def\Fxk{{\cal F}(x,k)}

\def\Feqpmxk{{\cal F}^{\pm}_{\rm eq}(x,k)}

\def\Pxk{{\cal P}(x,k)}

\def\Peqpmxk{{\cal P}^{\pm}_{\rm eq}(x,k)}

\title{Thermodynamic versus kinetic approach to polarization-vorticity coupling}

\author[a,b]{Wojciech Florkowski,}
\author[a]{Avdhesh Kumar,}
\author[a]{Radoslaw Ryblewski}

\affiliation[a]{Institute of Nuclear Physics Polish Academy of Sciences,\\ PL-31342 Krakow, Poland}
\affiliation[b]{Jan Kochanowski University, \\PL-25406 Kielce, Poland}

\emailAdd{wojciech.florkowski@ifj.edu.pl}
\emailAdd{avdhesh.kumar@ifj.edu.pl}
\emailAdd{radoslaw.ryblewski@ifj.edu.pl}

\abstract{We critically compare thermodynamic and kinetic approaches, that have been recently used to study relations between the spin polarization and fluid vorticity in systems consisting of spin-one-half particles. The thermodynamic approach refers to general properties of global thermal equilibrium with a rigid-like rotation and demonstrates that the spin-polarization and thermal-vorticity tensors are equal. On the other hand, the kinetic approach uses the concept of the Wigner function and its semiclassical expansion. In most of the works done so far, the Wigner functions satisfy kinetic equations with a vanishing collision term.  We show that this assumption restricts significantly applicability of such frameworks and, in contrast to many claims found in the literature, does not allow for drawing any conclusions regarding the relation between the thermal-vorticity and spin-polarization tensors, except for the fact that the two should be constant in global equilibrium.  We further show how the kinetic-theory equations including spin degrees of freedom can be used to formulate a hydrodynamic framework for particles with spin. We define hydrodynamic equations starting separately from the formulation by de~Groot, van~Leeuwen, and van~Weert and from the canonical formalism. In the former case the energy-momentum tensor is symmetric and the spin tensor is conserved, while in the later  case the energy-momentum tensor is not symmetric and the spin tensor is not conserved. Nevertheless, in the two cases the total angular momentum is always conserved. Interestingly, the two approaches are connected by the pseudo-gauge transformation, which we explicitly define.}

\begin{document} 
\maketitle
\flushbottom

\section{Introduction}
\label{sec:intro}

Recently, in connection with the first positive measurements of the $\Lambda$--hyperon spin polarization~\cite{STAR:2017ckg,Adam:2018ivw}, a lot of interest has been triggered in theoretical studies analyzing the spin polarization and vorticity formation in heavy-ion collisions. One expects that the spin polarization can be related to the global rotation of the strongly interacting matter created in the non-central collisions, in a~way similar to the magnetomechanical Barnett effect \cite{Barnett:1935} and Einstein and de Haas effect \cite{dehaas:1915}. Vorticity can also give rise to new phenomena such as the chiral vortical effect \cite{Kharzeev:2010gr, Kharzeev:2015znc}. Interestingly, the longitudinal polarization of ${\bar \Lambda}$ was discussed already in 1980s by Jacob and Rafelski in connection with the quark-gluon plasma formation \cite{Jacob:1987sj}. However,  the negative results were reported by the first heavy-ion experiments that measured the $\Lambda$ spin polarization in Dubna \cite{Anikina:1984cu}, at CERN  \cite{Bartke:1990cn} and BNL \cite{Abelev:2007zk}.

In the context of various effects associated with the spin polarization and vorticity, many theoretical studies have been performed that refer to the spin-orbit coupling \cite{Liang:2004ph,Liang:2004xn,Gao:2007bc,Chen:2008wh}, statistical properties of matter in equilibrium \cite{Weert:1982,Zubarev:1979,Becattini:2009wh,Becattini:2012tc,Becattini:2013fla,Becattini:2015nva,Hayata:2015lga}, and kinetic models with spin \cite{Gao:2012ix,Chen:2012ca,Fang:2016vpj,Fang:2016uds}. Moreover, closely related works on hydrodynamics with triangle anomalies \cite{Son:2009tf,Kharzeev:2010gr} and on the Lagrangian formulation of hydrodynamics have been reported in Refs.~\cite{Montenegro:2017rbu,Montenegro:2017lvf,Montenegro:2018bcf}.

A natural framework for dealing simultaneously with polarization and vorticity would be relativistic hydrodynamics of polarized fluids. An example of such a framework has been recently proposed in Refs.~\cite{Florkowski:2017ruc,Florkowski:2017dyn}. It is based on the local equilibrium distribution functions for particles and antiparticles with spin $\onehalf$, in the form introduced in~Ref.~\cite{Becattini:2013fla}. This framework can describe the full space-time evolution of the spin polarization in systems created in high-energy nuclear collisions. We note, that the inclusion of the spin degrees of freedom into a hydrodynamic approach represents one of several novel developments in relativistic hydrodynamics which forms the basis for our understanding of space-time evolution of matter created in heavy-ion collisions (for recent reviews on progress in relativistic hydrodynamics see \cite{Florkowski:2017olj,Romatschke:2017ejr}).

\medskip
In this paper we perform a detailed comparison of the thermodynamic and kinetic approaches which deal with the phenomenon of polarization-vorticity coupling in heavy-ion collisions. By the thermodynamic approach we mean a series of papers by Becattini and his collaborators~\cite{Becattini:2009wh,Becattini:2012tc,Becattini:2013fla,Becattini:2013vja,Becattini:2015nva,Becattini:2016gvu,Becattini:2017gcx}, where the authors analyze predominantly the properties of matter in global equilibrium with a rigid rotation. On the other hand, by the kinetic approach we mean here Refs.~\cite{Gao:2012ix,Chen:2012ca,Fang:2016uds,Fang:2016vpj}, where collisionless kinetic equations for the Wigner functions of spin-$\onehalf$ particles have been studied. 

Similarly to Refs.~\cite{Gao:2012ix,Chen:2012ca,Fang:2016uds,Fang:2016vpj} we perform herein a semiclassical expansion of the Wigner function. This method was successfully used in the past (see, for example, Refs. ~\cite{Elze:1986qd,Vasak:1987um,Elze:1989un,Zhuang:1995pd,Florkowski:1995ei}) to construct a classical limit of quantum kinetic equations, which yields dynamic equations for both the phase-space distribution functions and the spin phase-space densities. The novel feature of our present work is that we use the form of the equilibrium functions for particles with spin~$\onehalf$, proposed in Ref.~\cite{Becattini:2013fla}, as an input for the semiclassical expansion. In this way, we can check directly how the thermodynamic and kinetic frameworks are complementary to each other and what one approach implies for the other one. 

In order to make our formalism as simple as possible, and to concentrate primarily on the relation between the spin polarization and vorticity, we neglect in this work the effects of the electromagnetic and other mean fields. The inclusion of such fields is left for a separate analysis.

One of our findings is that recent formulations of the kinetic theory ~\cite{Gao:2012ix,Chen:2012ca,Fang:2016uds,Fang:2016vpj} do not imply the spin polarization induction by the vorticity. Although there exist solutions of the kinetic equations where the two phenomena are interconnected, they are in general independent. This is due to the fact that the collision term is neglected in such frameworks and the collisionless kinetic equation alone cannot imply the growth of polarization due to vorticity~\footnote{We do not discuss here the chiral kinetic theory \cite{Stephanov:2012ki,Chen:2014cla,Gorbar:2017toh} as its relation to the thermodynamic approach of Refs.~\cite{Becattini:2009wh,Becattini:2012tc,Becattini:2013fla} is not  known at the moment and requires a separate analysis.}. 

We further show that the kinetic-theory results demonstrating relations between polarization and vorticity correspond to the exact solutions of the collisionless kinetic equation. Thus, they can be interpreted as description of global thermodynamic equilibrium. Only in this case, the thermodynamic and kinetic results are fully consistent. To clarify this point, besides the concept of global and local equilibrium, we introduce also the ideas of extended global and extended local equilibrium.

Finally, we analyze different possible ways leading from the kinetic theory to the hydrodynamic equations with spin. They are all based on the application of the conservation laws for charge, energy, linear momentum, and angular momentum. Using the semiclassical expansion for the Wigner function, we introduce hydrodynamic equations starting from the kinetic-theory formulation by de Groot, van Leeuwen, and van Weert (GLW) \cite{deGroot:1980},  and using directly the canonical formalism \cite{Itzykson:1980rh}. In the GLW case the energy-momentum tensor is symmetric and the spin tensor is conserved, while in the canonical case the energy-momentum tensor is asymmetric and the spin tensor is not conserved (in both cases the total angular momentum is always conserved). Interestingly, the two approaches are connected by the pseudo-gauge transformation, which we have explicitly constructed.

\smallskip
{\it Conventions and notation:} Below we use the following conventions and notation for the metric tensor, Levi-Civita's tensor, and the scalar product: $g_{\mu\nu} =  \hbox{diag}(+1,-1,-1,-1)$, $\epsilon^{0123} = -\epsilon_{0123} = 1$, $a \cdot b = g_{\mu \nu} a^\mu b^\nu = a^0 b^0 - \av \cdot \bv$. Throughout the text we use $c = \hbar = k_B~=1$, however, we explicitly display $\hbar$ in the discussion of the semiclassical expansion of the Wigner function. All calculations are done using the Dirac representation for the gamma matrices. The operator $\Delta^{\mu\nu}$ projecting on the space orthogonal to the flow vector $u^\mu$ is defined as $\Delta^{\mu\nu} = g^{\mu\nu} - u^\mu u^\nu$.

The Lorentz invariant measure in the momentum space is denoted as $dP$, namely
\bel{eq:dP}
dP = \frac{d^3p}{(2 \pi )^3 E_p},
\eel
where $E_p = \sqrt{m^2 + \pv^2}$ is the on-mass-shell particle energy, and $p^\mu = (E_p, \pv)$. The particle momenta which are not necessarily on the mass shell and appear as arguments of the Wigner functions are denoted by the four-vector $k^\mu = (k^0, \kv)$.

The square brackets denote antisymmetrization, $t^{[\mu \nu]} =   \left(t^{\mu\nu} - t^{\nu\mu} \right)/2$. The symbol of tilde is used to denote dual tensors, which are obtained from the rank-two antisymmetric tensors by contraction with the Levi-Civita symbol and division by a factor of two. For example, $\omnLD$ denotes the dual spin polarization tensor defined by the equation
\bel{eq:dual}
\omnLD = \f{1}{2} \epsLmnab  \oabU,
\eel
where $\oabU$ is the original spin polarization tensor. The inverse transformation is
\bel{eq:dualdual}
\omega^{\rho \sigma} = -\f{1}{2} \epsilon^{\rho \sigma \mu \nu}
{\tilde \omega}_{\mu \nu}.
\eel
%

\section{Basic concepts and methodology}
\label{sec:bconcepts}

\subsection{Spinless particles --- global and local equilibrium}
\label{sec:spinless}

Before we start our discussion of various effects connected with spin, it is useful to recall basic features of the kinetic theory for spinless particles: In this case, the relativistic Boltzmann equation for the phase-space distribution function $f(x,p)$ contains two terms: the drift term and the collision integral. This can be schematically written as
\beq
p^\mu \p_\mu f(x,p) = C[f(x,p)].
\label{eq:simpkeq}
\eeq
The collision integral $C[f]$ vanishes in two special cases: i) for non-interacting, free streaming particles, and ii) for global or local thermodynamic equilibrium. In the first case the distribution function satisfies exactly the drift equation ($p^\mu \p_\mu f_{\rm fs}(x,p) = 0$) describing, unrelated to the present study,  free motion of particles. In the second case, which is of main interest for us, we have to distinguish between the global and local equilibrium. 

In the global thermodynamic equilibrium, the equilibrium distribution function $f_{\rm eq}(x,p)$ satisfies again the equation of the form $p^\mu \p_\mu f_{\rm eq}(x,p) = 0$, which leads in this case to the constraints on the hydrodynamic parameters used to specify the form of $f_{\rm eq}(x,p)$. In particular, the $\beta_\mu(x)$ field, defined traditionally as the ratio of the local fluid four-velocity $u_\mu(x)$ to the local temperature $T(x)$, satisfies the Killing equation
\beq
\p_\mu \beta_\nu(x) + \p_\nu \beta_\mu(x) = 0.
\label{eq:Killing}
\eeq
Equation~\rfn{eq:Killing} has the solution of the form~\footnote{The method of solving the Killing equation is presented in App.~\ref{sec:Killing}.}
\beq
\beta_\mu(x) =  \beta^0_\mu + \varpi^0_{\mu \nu} x^\nu,
\label{eq:Killingsol}
\eeq
where the vector $\beta^0_\mu$ and the antisymmetric tensor $\varpi^0_{\mu \nu}$ are constant. For any form of the field $\beta_\mu(x)$, we define thermal vorticity as the rotation 
\bel{eq:thvor}
\varpi_{\mu \nu} = -\frac{1}{2} \left(\p_\mu \beta_\nu - \p_\nu \beta_\mu \right).
\eel
Hence, Eqs.~\rfn{eq:Killing} and \rfn{eq:Killingsol} imply that the thermal vorticity in global equilibrium is constant, $\varpi_{\mu \nu}=\varpi^0_{\mu \nu}$. Additionally, in global equilibrium the ratio of the chemical potential to the local temperature should be constant, $\xi(x) = \mu(x)/T(x) = \xi^0 = \hbox{const}$.

In the case of local equilibrium, the right-hand side of~\rf{eq:simpkeq} vanishes, while its left-hand side, strictly speaking, does not. In this case one should add a correction $\delta f$ to the equilibrium function $f_{\rm eq}$, which describes dissipative phenomena. Nevertheless, if the gradients of local hydrodynamic variables are sufficiently small, the dissipative terms can be neglected. In this case the hydrodynamic variables in $f_{\rm eq}$ remain unconstrained. In order to determine them, one adds further assumptions, most commonly, that specific moments of~\rf{eq:simpkeq} in the momentum space (those that yield the conservation laws for energy, momentum or charge) vanish. This methodology leads to the perfect-fluid description.

\subsection{Particles with spin}
\label{sec:spin}

The treatment of the collisionless kinetic equation for the Wigner function ${\cal W}(x,k)$ that includes spin degrees of freedom has many features in common with the simple spinless system discussed above. As the free-streaming case is not interesting, we are left again with essentially two different physics cases which represent global and local thermodynamic equilibrium. Both of them can be analyzed with the help of the equilibrium distribution functions $f^+(x,p)$ and $f^-(x,p)$, for particles and antiparticles with spin $\onehalf$, introduced by Becattini and collaborators in \cite{Becattini:2013fla}. As the matter of fact, these functions are two-by-two Hermitian matrices that can be interpreted as spin density matrices for each value of the space-time position $x$ and momentum $p$. Besides typical dependence on the hydrodynamic variables $\beta_\mu = u_\mu/T$ and $\xi = \beta \mu = \mu/T$, they depend in addition on the antisymmetric spin polarization tensor $\omega_{\mu\nu}$ ($\omega_{\mu\nu} = -\omega_{\nu\mu}$). The equilibrium Wigner function $\Weqxk$, constructed from the functions $f^+(x,p)$ and $f^-(x,p)$, also depends on $\beta_\mu$, $\xi$, and $\omega_{\mu\nu}$. Consequently, it turns out that we can  distinguish between four rather than two different types of equilibrium. They can be classified as follows:
\begin{itemize}
	\item{} global equilibrium --- in this case the $\beta_\mu$ field is a Killing vector satisfying \rf{eq:Killing}, $\varpi_{\mu \nu} = -\frac{1}{2} \left(\p_\mu \beta_\nu - \p_\nu \beta_\mu \right) = \hbox{const}$, the spin polarization tensor is constant and agrees with thermal vorticity, $\omega_{\mu\nu} = \varpi_{\mu \nu}$, in addition $\xi = \hbox{const}$,
	\item{} extended global equilibrium --- $\beta_\mu$ field is a Killing vector, $\varpi_{\mu \nu} = -\frac{1}{2} \left(\p_\mu \beta_\nu - \p_\nu \beta_\mu \right) = \hbox{const}$, the spin polarization tensor is constant but $\omega_{\mu\nu} \neq \varpi_{\mu \nu}$, $\xi = \hbox{const}$, 
	\item{} local equilibrium --- $\beta_\mu$ field is not a Killing vector but we still have $\omega_{\mu\nu}(x) = \varpi_{\mu \nu}(x)$, $\xi$ is allowed to depend on space-time coordinates, $\xi = \xi(x)$,
	\item{} extended local equilibrium --- $\beta_\mu$ field is not a Killing vector and $\omega_{\mu\nu}(x) \neq \varpi_{\mu \nu}(x)$, moreover $\xi = \xi(x)$.
\end{itemize}

The global and extended global equilibrium states correspond to the case where $\Weqxk$ satisfies exactly the collisionless kinetic equations. On the other hand, in the local and extended local equilibrium states only certain moments of the kinetic equation for $\Weqxk$ can be set equal to zero. They can be used to construct perfect-fluid hydrodynamic equations including spin. 

We stress that in this work we assume that the collision term vanishes for each type of equilibrium listed above, provided the equilibrium Wigner function $\Weqxk$ has the form derived from the functions $f^+(x,p)$ and $f^-(x,p)$. This assumption should be verified in the future by detailed studies of various collision terms for particles with spin. Such studies may also shed new light on the form of the equilibrium distributions. Before the results of such investigations are known, we continue to assume that the collision term vanishes for $\Weqxk$.

Before we turn to discussion of the kinetic equation for the Wigner function $\Weqxk$ it is useful to characterize global thermodynamic equilibrium in the framework of relativistic quantum mechanics. This leads to a natural distinction between the global and extended global equilibrium.

\section{Global thermodynamic equilibrium in relativistic \\ quantum mechanics} \label{sec:global}
In this section we introduce general features of global thermodynamic equilibrium constructed in the framework of relativistic quantum mechanics. We follow here closely the treatment of Zubarev \cite{Zubarev:1974} and Becattini \cite{Becattini:2012tc}. The main object of interest in this approach is a density operator ${\hat \rho}$ defined by the expression
\bel{eq:rho}
{\hat \rho}(t) = \exp\left[-\int d^3\Sigma_\mu(x) \left( {\hat T}^{\mu\nu}(x) b_\nu(x) 
- \f{1}{2} {\hat J}^{\mu, \alpha\beta}(x) \omega_{\alpha \beta}(x) - {\hat{N}^{\mu}}(x)  \xi(x) \right)\right].
\eel
Here $d^3\Sigma_\mu$ is an element of a space-like, three-dimensional hypersurface $\Sigma_\mu$. We may assume that it corresponds to a fixed value of the time coordinate. In this case $d^3\Sigma_\mu=(dV,0,0,0)$ and ${\hat \rho}$ becomes a function of $t$. The operators ${\hat T}^{\mu\nu}(x)$, ${\hat J}^{\mu, \alpha\beta}(x)$ and ${\hat{N}^{\mu}}(x)$ are quantum versions of the energy-momentum tensor, angular momentum tensor, and charge current. They obey the following conservation laws:
\beq
\p_\mu {\hat T}^{\mu\nu}(x)  = 0,  \label{cons_enm}
\eeq
\beq
\p_\mu {\hat J}^{\mu, \alpha\beta}(x)  = 0, \label{cons_angm}
\eeq
\beq
\p_\mu {\hat N}^{\mu}(x)  = 0. \label{cons_crt}
\eeq
Note that ${\hat J}^{\mu, \alpha\beta}(x)$ is asymmetric in the last two indices, ${\hat J}^{\mu, \alpha\beta}(x)=-{\hat J}^{\mu, \beta \alpha}(x)$ and can be, in general, represented as a sum of the orbital and spin parts 
\beq
{\hat J}^{\mu, \alpha\beta}(x) = {\hat  L}^{\mu, \alpha\beta}(x) + {\hat S}^{\mu, \alpha\beta}(x).
\label{eq:angular_momentum}
\eeq
The orbital part is expressed by the space-time coordinates and the energy-momentum-tensor components
\beq
{\hat L}^{\mu, \alpha\beta}(x)  = x^\alpha {\hat T}^{\mu \beta}(x) - x^\beta {\hat  T}^{\mu \alpha}. 
\label{eq:orbital_ang_mntm}
\eeq
Using \rftwo{cons_enm}{cons_angm} we find
\beq
\p_\mu {\hat S}^{\mu, \alpha\beta}(x) = {\hat T}^{\beta \alpha}(x) - {\hat T}^{\alpha \beta}(x). 
\label{eq:spin_ang_mntm}
\eeq
Thus, the spin contribution to the angular momentum is usually not conserved --- it is conserved only if the energy momentum operator ${\hat T}^{\alpha \beta}(x)$ is symmetric.  The functions $b_\nu(x)$, $\omega_{\alpha\beta}(x)$, and $\xi(x)$ are Lagrange multipliers that should be chosen to maximize the system's entropy. Note that $\omega_{\alpha\beta}(x) = - \omega_{\beta \alpha}(x)$ as any symmetric part of $\omega_{\alpha\beta}(x)$ does not give contribution to \rf{eq:rho}.

In global thermodynamic equilibrium we require that the operator ${\hat \rho}(t)$ is independent of time. This condition leads to the constraint 
\beq
&& \p_\mu \left({\hat T}^{\mu\nu}(x) b_\nu(x)  - \f{1}{2} {\hat J}^{\mu, \alpha\beta}(x) \omega_{\alpha \beta}(x) 
-{\hat{N}^{\mu}(x)} \xi(x) \right) \nn \\
&& \hspace{1cm}
= {\hat T}^{\mu\nu}(x) \left( \p_\mu b_\nu(x) \right) -
\f{1}{2} {\hat J}^{\mu, \alpha\beta}(x)  \left(  \p_\mu  \omega_{\alpha \beta}(x)\right) 
- {\hat{N}^{\mu}}(x) \p_\mu \xi(x)= 0. 
\label{div}
\eeq
From this equation we can conclude that the parameters $\xi$ and $\omega_{\alpha\beta}$ are constants, $\xi=\xi^0$ and $\omega_{\alpha\beta}=\omega^0_{\alpha\beta}$~\footnote{We note that if the tensor ${\hat J}^{\mu, \alpha\beta}$ has additional symmetries, for example, it is completely antisymmetric, more general solutions for $ \omega_{\alpha \beta}(x)$ may exist.}. The form of $b_{\nu}$ depends on the symmetry of the energy-momentum tensor ${\hat T}^{\mu\nu}(x)$. For symmetric ${\hat T}^{\mu\nu}$, we require that $\p_\mu b_\nu + \p_\nu b_\mu =0$, hence $b_\nu$ is a Killing vector, 
\bel{eq:bS}
b_{\nu} = b^{0}_{\nu}+\delta\omega^{0}_{\nu\rho}\, x^{\rho},  
\eel
where $b^{0}_{\nu}$ and $\delta\omega^{0}_{\nu\rho}=-\delta\omega^{0}_{\rho\nu}$ are constants. On the other hand, for a not symmetric (asymmetric) ${\hat T}^{\mu\nu}$ we require that  $\p_\mu b_\nu=0$, hence $b_\nu$ must be a constant vector, $b_{\nu} = b^{0}_{\nu}$.

Using the decomposition of the angular momentum into the orbital and spin parts, see \rf{eq:angular_momentum}, one can show that the two cases discussed above can be expressed by a single form of the density operator 
\beq
{\hat \rho}_{\rm EQ}&=& \exp\left[-\int d^3\Sigma_\mu(x) \left( {\hat T}^{\mu\nu}(x)\beta_\nu(x)
- \f{1}{2} {\hat S}^{\mu, \alpha\beta}(x) \omega^0_{\alpha \beta}-{\hat{N}^{\mu}}(x) \xi^0  \right) \right]. 
\label{eq:rhoEQ1}
\eeq
For asymmetric energy-momentum tensor $\beta_\mu(x) = b^0_\mu + \omega^0_{\mu \gamma} x^\gamma$ (with constant $b^0_\mu$ and $\omega^0_{\mu \gamma}$). This implies that $\beta_\mu(x)$ is a Killing vector and thermal vorticity defined by \rf{eq:thvor} agrees with the spin polarization tensor $\omega_{\mu \gamma}=\omega^0_{\mu \gamma}$. On the other hand, for symmetric energy-momentum tensor $\beta_\mu(x) = b^0_\mu + (\delta\omega^0_{\mu \gamma} + \omega^0_{\mu \gamma} ) x^\gamma$ (with constant $b^0_\mu$, $\delta\omega^0_{\mu \gamma}$ and $\omega^0_{\mu \gamma}$). In this case $\beta_\mu(x)$ is again a Killing vector, however, thermal vorticity defined by \rf{eq:thvor} does not necessarily agree with the spin polarization tensor. 

Our discussion indicates that depending on the symmetry of the energy-momentum tensor, we may deal with global or extended global equilibrium, as they have been defined in the end of Sec.~\ref{sec:bconcepts}. For completeness, we define the statistical operator for local equilibrium by the same form as \rf{eq:rhoEQ1},
\beq
{\hat \rho}_{\rm eq}&=& \exp\left[-\int d^3\Sigma_\mu(x) \left( {\hat T}^{\mu\nu}(x)\beta_\nu(x)
- \f{1}{2} {\hat S}^{\mu, \alpha\beta}(x) \omega_{\alpha \beta}(x)-{\hat{N}^{\mu}} (x) \xi(x) \right) \right],
\label{eq:rhoEQ2}
\eeq
allowing for arbitrary form of $\beta_\mu(x)$ and $\xi(x)$, and for two options for $\varpi_{\mu\nu}(x)$: either $\varpi_{\mu\nu}(x) =  \omega_{\mu\nu}(x)$ (local equilibrium) or $\varpi_{\mu\nu}(x) \neq  \omega_{\mu\nu}(x)$ (extended local equilibrium). 

\section{Equilibrium Wigner functions} 
\label{sec:geq}

\subsection{Spin-dependent equilibrium distribution functions} \label{sec:spindistr}

To include the spin degrees of freedom, the scalar equilibrium distribution functions are generalized to two-by-two spin density matrices for each value of the space-time position $x$ and momentum $p$~\cite{deGroot:1980},
\beq
\left[ f^+(x,p) \right]_{rs}  \equiv  \fplusrsxp &=& 
\frac{1}{2m} \ubarrp X^+ \usp, \label{fplusrsxp}  \\
\left[ f^-(x,p) \right]_{rs}  \equiv \fminusrsxp &=& 
- \frac{1}{2m}\vbarsp X^- \vrp. \label{fminusrsxp}
\eeq
Here $m$ is the (anti)particle mass, while $\urp$ and $\vrp$ are Dirac bispinors (with the spin indices $r$ and $s$ running from 1~to~2), and the normalizations:
\bel{eq:unorm}
\ubarrp \usp=\,2m\, \delta_{rs}, \qquad 
\sum_{r=1}^{2} u^r_\alpha(p) {\bar u}^r_\beta(p) = 
(\slashed{p}+m)_{\alpha \beta},
\eel
\bel{eq:vnorm}
\vbarrp \vsp=-\,2m\, \delta_{rs}, \qquad 
\sum_{r=1}^{2} v^r_\alpha(p) {\bar v}^r_\beta(p) = 
(\slashed{p}-m)_{\alpha \beta}.
\eel
Note the minus sign and different ordering of spin indices in \rf{fminusrsxp} compared to \rf{fplusrsxp}. The objects $f^\pm(x,p)$ are two-by-two Hermitian matrices with the matrix elements defined by \rftwo{fplusrsxp}{fminusrsxp}.

Following Ref.~\cite{Becattini:2013fla}, we use the four-by-four matrices
\bel{XpmM}
X^{\pm} =  \exp\left[\pm \xi(x) - \bmu(x) \pmu \right] M^\pm, 
\eel
where
\bel{Mpm}
M^\pm = \exp\left[ \pm \f{1}{2} \omnL(x)  \SmunuU \right] .
\eel
In \rftwo{XpmM}{Mpm} we use the same notation as that introduced in the previous sections, namely: $\beta^\mu(x)= \umU(x)/T(x)$ and $\xi(x) = \mu(x)/T(x)$, with $\mu(x)$ being the chemical potential (connected with a charge that can be identified, for example, with the baryon number or electric charge). The quantity $\omnL(x)$ is the spin polarization tensor, while  $\SmunuU$  is the spin operator expressed in terms of the Dirac gamma matrices, $\SmunuU = (i/4) [\gamma^\mu,\gamma^\nu]$. 

For the sake of simplicity, we restrict ourselves to classical Boltzmann statistics in this work. Following \rfc{Florkowski:2017ruc} we further assume that the spin polarization tensor $\omnL$ satisfies the two conditions~\footnote{The conditions \rfn{eq:conditions} are satisfied in a natural way if only space components $\omega_{ij}$ are different from zero. This happens, for example, in the case of global equilibrium with a rigid rotation. The non-zero $\omega_{0i}$ components appear, on the other hand, for global equilibrium with a constant acceleration along the fluid stream lines, see Refs.~\cite{Becattini:2015nva,Becattini:2017ljh,Florkowski:2018myy,Prokhorov:2018qhq,Prokhorov:2018bql}}.
\beq
\omnL \omnU \geq 0, \quad \omnL \omnUD = 0,
\label{eq:conditions}
\eeq
In this case we introduce the variables $\zeta$ and $\Omega$ defined by the expression
\bel{eq:zeta}
\zeta = \f{\Omega}{T} = \f{1}{2} \sqrt{ \frac{1}{2} \omnL \omnU }.
\eel
It turns out, see \rfc{Florkowski:2017ruc}, that $\Omega$ plays a role of the chemical potential related with spin. Using \rf{eq:conditions} one finds
\bel{eq:Mpmexp}
M^\pm &=& \cosh(\zeta) \pm  \f{\sinh(\zeta)}{2\zeta}  \, \omnL \SmunuU.
\eel

\subsection{Equilibrium Wigner functions} \label{sec:eqWig}

The equilibrium phase-space distribution functions $f^+(x,p)$ and $f^-(x,p)$ can be used to determine explicit expressions for the corresponding equilibrium (particle and antiparticle) Wigner functions. We construct them using the expressions from Ref.~\cite{deGroot:1980},
\beq
\Weqpxk = \frac{1}{2} \sum_{r,s=1}^2 \int dP\,
\delta^{(4)}(k-p) u^r(p) {\bar u}^s(p) f^+_{rs}(x,p),
\label{eq:Weqpxk}
\eeq
\beq
\Weqmxk = -\frac{1}{2} \sum_{r,s=1}^2 \int dP\,
\delta^{(4)}(k+p) v^s(p) {\bar v}^r(p) f^-_{rs}(x,p).
\label{eq:Weqmxk}
\eeq
The total Wigner function is a simple sum of these two contributions
\bel{eq:totW}
\Weqxk = \Weqpxk + \Weqmxk.
\eel
Using Eqs.~\rfn{fplusrsxp}--\rfn{eq:vnorm} we find
\beq
\Weqpxk = \frac{1}{4 m}  \int dP\,
\delta^{(4)}(k-p) (\slashed{p}+m) X^+ (\slashed{p}+m),
\label{eq:Weqpxk1}
\eeq
\beq
\Weqmxk = \frac{1}{4 m}  \int dP\,
\delta^{(4)}(k+p) (\slashed{p}-m) X^- (\slashed{p}-m).
\label{eq:Weqmxk1}
\eeq
With the help of \rf{eq:Mpmexp} we can further rewrite these equations in the following form
\beq
\Weqpxk &=& \frac{e^\xi}{4 m}  \int dP
\,e^{-\beta \cdot p }\,\, \delta^{(4)}(k-p) \nn \\
&& \times  \left[2m (m+\slashed{p}) \cosh(\zeta)+  \f{\sinh(\zeta)}{2\zeta}  \, \omnL \,(\slashed{p}+m) \SmunuU (\slashed{p}+m) \right],
\label{eq:Weqpxk2}
\eeq
\beq
\Weqmxk &=& \frac{e^{-\xi}}{4 m}  \int dP\,e^{-\beta \cdot p }\,\, \delta^{(4)}(k+p) \nn \\
&& \times  \left[2m (m-\slashed{p}) \cosh(\zeta)-  \f{\sinh(\zeta)}{2\zeta}  \, \omnL \,(\slashed{p}-m) \SmunuU (\slashed{p}-m) \right].
\label{eq:Weqmxk2}
\eeq

\section{Spinor decomposition of the equilibrium Wigner function} \label{sec:spdecWeq}

\subsection{Clifford-algebra expansion} \label{sec:clifford}

The equilibrium Wigner functions $\Weqpmxk$, being four-by-four matrices satisfying the relations $\Weqpmxk = \gamma_0 \Weqpmxk^\dagger \gamma_0$, can be always expanded in terms of the 16 independent generators of the Clifford algebra \cite{Itzykson:1980rh,Vasak:1987um},
\beq
\Weqpmxk &=& \f{1}{4} \left[ \Feqpmxk + i \gamma_5 \Peqpmxk + \gamma^\mu {\cal V}^\pm_{{\rm eq}, \mu}(x,k) \right. \nn \\
&& \left.  \hspace{1cm} + \gamma_5 \gamma^\mu {\cal A}^\pm_{{\rm eq}, \mu}(x,k)
+ \SmnU {\cal S}^\pm_{{\rm eq}, \mu \nu}(x,k) \right].
\label{eq:wig_expansion}
\eeq
The coefficient functions in the equilibrium Wigner function expansion \rfn{eq:wig_expansion} can be obtained by the folowing traces:
\beq
\Feqpmxk&=&\tr\left[\Weqpmxk\right],
\label{eq:Feqpm}\\
\Peqpmxk&=&-i\,\tr\left[\gamma^5\Weqpmxk\right],
\label{eq:Peqpm} \\
{\cal V}^{\pm}_{{\rm eq}, \mu}(x,k)&=&\tr\left[\gamma_{\mu}\Weqpmxk\right],
\label{eq:Veqpm}\\
{\cal A}^\pm_{{\rm eq}, \mu}(x,k) &=& \tr\left[\gamma_{\mu}\gamma^5 \Weqpmxk\right],
\label{eq:Aeqpm}\\
{\cal S}^\pm_{{\rm eq}, \mu \nu}(x,k)&=&2\,\tr\left[\SmnL \Weqpmxk\right].
\label{eq:Seqpm}
\eeq
Using \rftwo{eq:Weqpxk2}{eq:Weqmxk2} in the expressions \rfn{eq:Feqpm}--\rfn{eq:Seqpm}, and employing the identities for the Dirac matrices \rfn{eq:idf}--\rfn{eq:ids},  see App. \ref{sec:trgammas}, we find
\beq
\Feqpmxk &=& 2 m \cosh(\zeta)\,\int dP\, \,e^{-\beta \cdot p \pm \xi}\,\,\delta^{(4)} (k\mp p),  \label{eq:FEeqpm} \\
\Peqpmxk &=& 0, \label{eq:PEeqpm} \\
{\cal V}^{\pm}_{{\rm eq}, \mu}(x,k) &=& \pm\,2 \cosh(\zeta) \, \int dP\,e^{-\beta \cdot p \pm \xi}\,\,\delta^{(4)} (k\mp p)\,p_{\mu}, \label{eq:VEeqpm} \\
{\cal A}^\pm_{{\rm eq}, \mu}(x,k) &=& -\frac{\sinh(\zeta)\, }{\zeta} \,\int dP\,e^{-\beta \cdot p \pm \xi}\,\,\delta^{(4)}(k\mp p)\, \tilde{\omega }_{\mu \nu}\,p^{\nu},
\label{eq:AEeqpm} \\
{\cal S}^\pm_{{\rm eq}, \mu \nu}(x,k) &=& \! \pm\frac{ \sinh(\zeta) }{m \zeta} \!\int \!dP\,e^{-\beta \cdot p \pm\xi}\,\,\delta^{(4)}(k\mp p) 
\left[  \left( p_\mu \omega_{\nu \alpha} -  p_\nu \omega_{\mu \alpha} \right) p^\alpha \!+\! m^2\omega_{\mu \nu} \right]\!.\,\,\,\,\,\,\,
\label{eq:SEeqpm} 
\eeq
%

\subsection{Relations between equilibrium coefficient functions} \label{sec:relations}

Using Eqs.~\rfn{eq:FEeqpm}--\rfn{eq:SEeqpm} one can verify that the equilibrium coefficient functions satisfy the following set of constraints:
\bel{eq:Wid1}
k^\mu \, {\cal V}^{\pm}_{{\rm eq}, \mu}(x,k) = 
m \, {\cal F}^{\pm}_{{\rm eq}}(x,k),
\eel
\bel{eq:Wid2}
k_\mu \, {\cal F}^{\pm}_{{\rm eq}}(x,k) = m \, {\cal V}^{\pm}_{{\rm eq}, \mu}(x,k),
\eel
\bel{eq:Wid3}
{\cal P}^\pm_{{\rm eq}}(x,k) = 0,
\eel
\bel{eq:Wid4}
k^\mu \, {\cal A}^{\pm}_{{\rm eq}, \,\mu}(x,k) = 0,
\eel
\bel{eq:Wid5}
k^\mu \, {\cal S}^{\pm}_{{\rm eq}, \,\mu \nu}(x,k) = 0.
\eel
\bel{eq:Wid6}
k^\beta \, {\tilde {\cal S}}^{\pm}_{{\rm eq}, \mu \beta}(x,k) + m \, {\cal A}^{\pm}_{{\rm eq}, \,\mu}(x,k) = 0,
\eel
\bel{eq:Wid7}
\epsilon_{\mu \nu \alpha \beta} \, k^\alpha \, {\cal A}^{\pm \, \beta}_{{\rm eq}}(x,k) + m \, {\cal S}^{\pm}_{{\rm eq}, \,\mu \nu}(x,k) = 0.
\eel
We note that such constraints are fulfilled also by the total Wigner function given by the sum of the particle and antiparticle contributions, see \rf{eq:totW}. We also note that Eqs.~\rfn{eq:Wid1}--\rfn{eq:Wid7} follow from the algebraic structure of the equilibrium Wigner functions and are satsified for any form of the fields: $\beta_\mu(x)$, $\xi(x)$, and $\omega_{\mu \nu}(x)$. Thus, they hold for four different types of equilibrium specified in the end of Sec.~\ref{sec:bconcepts}.

\section{Semi-classical expansion} \label{sec:semiclass}

In the previous section we have introduced the spinor decomposition of the equilibrium Wigner functions and obtained explicit expressions for the equilibrium coefficient functions. Such a decomposition can be naturally used for any Wigner function (describing particles with spin $\onehalf$) and, in fact, it was frequently used in the past to derive classical kinetic equations from the underlying quantum field theory~\cite{Elze:1986qd,Vasak:1987um,Elze:1989un,Zhuang:1995pd,Florkowski:1995ei}). In this section we follow closely this approach and write
\beq
\Wxk &=& \f{1}{4} \left[ \Fxk + i \gamma_5 \Pxk + \gamma^\mu {\cal V}_{\mu}(x,k) \right. \nn \\
&& \left.  \hspace{1cm} + \gamma_5 \gamma^\mu 
{\cal A}_{\mu}(x,k)
+ \SmnU {\cal S}_{\mu \nu}(x,k) \right].
\label{eq:gen_wig_expansion}
\eeq
In the case where the effects of both the mean fields and collisions can be neglected, the Wigner function satisfies the equation of the form
\bel{eq:eqforW}
\left(\gamma_\mu K^\mu - m \right) {\cal W}(x,k) = 0.
\eel
Here $K^\mu$ is the operator defined by the expression
\bel{eq:K}
K^\mu = k^\mu + \frac{i \hbar}{2} \,\p^\mu.
\eel
Using \rftwo{eq:gen_wig_expansion}{eq:K} in \rf{eq:eqforW} and comparing the real and imaginary parts of the coefficients in the Clifford-algebra basis we obtain two sets of equations. The real parts give:
\beq
k^\mu {\cal V}_\mu - m {\cal F} &=& 0,
\label{eq:rF} \\
\frac{\hbar}{2} \p^\mu {\cal A}_\mu + m {\cal P} &=& 0,
\label{eq:rP} \\
k_\mu {\cal F} - \frac{\hbar}{2} \p^\nu {\cal S}_{\nu\mu}
- m {\cal V}_\mu &=& 0,
\label{eq:rV} \\
-\frac{\hbar}{2} \p_\mu {\cal P} + 
k^\beta {\tilde {\cal S}}_{\mu \beta} + m {\cal A}_\mu &=& 0,
\label{eq:rA} \\
\frac{\hbar}{2} \left( \p_\mu {\cal V}_\nu - \p_\nu {\cal V}_\mu \right)
- \epsilon_{\mu \nu \alpha \beta} k^\alpha {\cal A}^\beta - m {\cal S}_{\mu \nu} &=& 0,
\label{eq:rS}
\eeq
while the imaginary parts yield:
\beq
\hbar \p^\mu {\cal V}_\mu &=& 0,
\label{eq:iF} \\
k^\mu {\cal A}_\mu &=& 0,
\label{eq:iP} \\
\frac{\hbar}{2} \p_\mu {\cal F} + k^\nu {\cal S}_{\nu\mu}
&=& 0,
\label{eq:iV} \\
k_\mu {\cal P} + \frac{\hbar}{2}  \p^\beta {\tilde {\cal S}}_{\mu \beta} &=& 0,
\label{eq:iA} \\
\left(k_\mu {\cal V}_\nu - k_\nu {\cal V}_\mu \right)
+\frac{\hbar}{2} \epsilon_{\mu \nu \alpha \beta} \p^\alpha {\cal A}^\beta &=& 0.
\label{eq:iS}
\eeq

The form of Eqs.~\rfn{eq:rF}--\rfn{eq:iS} suggests that we can search for solutions for the expansion coefficient functions in the form of the series:
\bel{eq:series1}
{\cal F} = {\cal F}^{(0)} + \hbar {\cal F}^{(1)} +  \hbar^2 {\cal F}^{(2)}+ \cdots, \quad 
{\cal P} = {\cal P}^{(0)} + \hbar {\cal P}^{(1)} +  \hbar^2 {\cal P}^{(2)} + \cdots,
\eel
\bel{eq:series2}
{\cal V}_\mu = {\cal V}^{(0)}_\mu + \hbar {\cal V}^{(1)}_\mu +  \hbar^2 {\cal V}^{(2)}_\mu + \cdots, \quad 
{\cal A}_\mu = {\cal A}^{(0)}_\mu + \hbar {\cal A}^{(1)}_\mu +  \hbar^2 {\cal A}^{(2)}_\mu + \cdots,
\eel
\bel{eq:series3}
{\cal S}_{\mu\nu} = {\cal S}^{(0)}_{\mu\nu} + \hbar {\cal S}^{(1)}_{\mu\nu} +  \hbar^2 {\cal S}^{(2)}_{\mu\nu} + \cdots. \quad 
\eel

\subsection{Zeroth order} \label{sec:zeroth}

The leading order (the zeroth order in $\hbar$) of the real parts gives:
\beq
k^\mu {\cal V}^{(0)}_\mu - m {\cal F}^{(0)} &=& 0,
\label{eq:rF0} \\
{\cal P}^{(0)} &=& 0,
\label{eq:rP0} \\
k_\mu {\cal F}^{(0)} - m {\cal V}^{(0)}_\mu &=& 0,
\label{eq:rV0} \\
k^\beta {\tilde {\cal S}}_{\mu \beta}^{(0)} + m {\cal A}^{(0)}_\mu &=& 0,
\label{eq:rA0} \\
\epsilon_{\mu \nu \alpha \beta} k^\alpha {\cal A}_{(0)}^\beta + m {\cal S}_{\mu \nu}^{(0)} &=& 0,
\label{eq:rS0}
\eeq
while the leading order of the imaginary parts gives~\footnote{The imaginary part of the scalar zeroth-order part of \rf{eq:eqforW} vanishes, see \rf{eq:iF}, whereas the imaginary part of the axial-vector zeroth-order part of \rf{eq:eqforW} gives \rf{eq:rP0}, see \rf{eq:iA}. Therefore, we consider only three equations obtained from the imaginary parts.}
\beq
k^\mu {\cal A}^{(0)}_\mu &=& 0,
\label{eq:iP0} \\
k^\nu {\cal S}^{(0)}_{\nu\mu}
&=& 0,
\label{eq:iV0} \\
k_\mu {\cal V}^{(0)}_\nu - k_\nu {\cal V}^{(0)}_\mu 
&=& 0.
\label{eq:iS0}
\eeq
Equations \rfn{eq:rF0}--\rfn{eq:iS0} indicate the coefficients ${\cal F}_{(0)}$ and ${\cal A}^\mu_{(0)}$ may be treated as the basic independent ones, provided ${\cal A}^\mu_{(0)}$ satisfies the orthogonality condition \rfn{eq:iP0}. The coefficient ${\cal V}^\mu_{(0)}$ is defined by \rf{eq:rV0}, which gives
\bel{eq:rV0a}
{\cal V}^\mu_{(0)} = \frac{k^\mu}{m} {\cal F}_{(0)},
\eel
and the coefficient ${\cal S}_{\mu \nu}^{(0)}$ is obtained from \rf{eq:rS0},
\bel{eq:rS0a}
{\cal S}_{\mu \nu}^{(0)} = -\frac{1}{m} \epsilon_{\mu\nu \alpha \beta} k^\alpha {\cal A}^\beta_{(0)}.
\eel
Equation \rfn{eq:rS0a} leads directly to the dual tensor ${\tilde {\cal S}}_{\mu \nu}^{(0)}$ of the form
\bel{eq:rS0ad}
{\tilde {\cal S}}_{\mu \nu}^{(0)} = \frac{1}{m} \left( 
k^\mu {\cal A}^\nu_{(0)} - k^\nu {\cal A}^\mu_{(0)} \right).
\eel
One can easily check that expressions~\rfn{eq:rV0a}--\rfn{eq:rS0ad} solve Eqs.~\rfn{eq:rF0}--\rfn{eq:rS0} and Eqs.~\rfn{eq:iP0}--\rfn{eq:iS0} if the axial-vector coefficient ${\cal A}^\mu_{(0)}$ fulfills \rf{eq:iP0}. 

\subsection{First order} \label{sec:first}

The next-to-leading order (the first order in $\hbar$) of the real parts gives:
\beq
k^\mu {\cal V}^{(1)}_\mu - m {\cal F}^{(1)} &=& 0,
\label{eq:rF1} \\
\frac{1}{2} \p^\mu  {\cal A}^{(0)}_\mu + m {\cal P}^{(1)} &=& 0,
\label{eq:rP1} \\
k_\mu {\cal F}^{(1)} - \frac{1}{2} \p^\nu {\cal S}^{(0)}_{\nu \mu}-
m {\cal V}^{(1)}_\mu &=& 0,
\label{eq:rV1} \\
-\frac{1}{2} \p_\mu {\cal P}_{(0)} +
k^\beta {\tilde {\cal S}}_{\mu \beta}^{(1)} + m {\cal A}^{(1)}_\mu &=& 0,
\label{eq:rA1} \\
\frac{1}{2} \left(\p_\mu {\cal V}^{(0)}_\nu - \p_\nu {\cal V}^{(0)}_\mu \right) - \epsilon_{\mu \nu \alpha \beta} k^\alpha {\cal A}_{(1)}^\beta - m {\cal S}_{\mu \nu}^{(1)} &=& 0.
\label{eq:rS1}
\eeq
Equation \rfn{eq:rP1} defines the first order contribution to the pseudoscalar coefficient
\bel{eq:rP1a}
{\cal P}^{(1)}  = -\frac{1}{2m} \, \p^\mu  {\cal A}^{(0)}_\mu. 
\eel
Similarly, \rf{eq:rV1} can be interpreted as the definition of the first-order vector coefficient
\bel{eq:rV1a}
{\cal V}^{(1)}_\mu &=& \frac{1}{m} \left(k_\mu {\cal F}^{(1)} 
- \frac{1}{2} \p^\nu {\cal S}^{(0)}_{\nu \mu} \right),
\eel
while \rf{eq:rS1} defines the first-order tensor coefficient
\bel{eq:rS1a}
{\cal S}_{\mu \nu}^{(1)} = \frac{1}{2m} \left(\p_\mu {\cal V}^{(0)}_\nu - \p_\nu {\cal V}^{(0)}_\mu \right) 
- \frac{1}{m} \epsilon_{\mu \nu \alpha \beta} k^\alpha {\cal A}_{(1)}^\beta.
\eel
By contraction of \rf{eq:rS1a} with the Levi-Civita tensor we find the dual first-order tensor coefficient 
\bel{eq:rS1ad}
{\tilde {\cal S}}_{\mu \nu}^{(1)} = \frac{1}{4m^2} 
\epsilon^{\mu \nu \alpha \beta}
\left(k_\alpha \p_\beta  - k_\beta \p_\alpha  \right) 
{\cal F}^{(0)} 
- \frac{1}{m} \epsilon_{\mu \nu \alpha \beta} k^\alpha {\cal A}_{(1)}^\beta.
\eel
Using \rf{eq:rS1ad} in \rf{eq:rA1} we find that the first-order axial coefficient should also be  orthogonal to $k^{\mu}$, namely $k_\mu {\cal A}_{(1)}^\mu = 0$.

The first order imaginary parts give:
\beq
\p^\mu {\cal V}^{(0)}_\mu &=& 0,
\label{eq:iF1} \\
k^\mu {\cal A}^{(1)}_\mu &=& 0,
\label{eq:iP1} \\
\frac{1}{2} \p_\mu {\cal F}^{(0)} + k^\nu {\cal S}^{(1)}_{\nu\mu}
&=& 0,
\label{eq:iV1} \\
k_\mu {\cal P}^{(1)} + \frac{1}{2} \p^\beta {\tilde {\cal S}}^{(0)}_{\mu\beta} &=& 0,
\label{eq:iA1} \\
k_\mu {\cal V}^{(1)}_\nu - k_\nu {\cal V}^{(1)}_\mu 
+ \frac{1}{2} \epsilon_{\mu\nu \alpha \beta} \, \p^\alpha {\cal A}_{(0)}^\beta
&=& 0.
\label{eq:iS1}
\eeq
Combining \rf{eq:iF1} with \rf{eq:rV0a} we find the important formula
\bel{eq:kineqF0}
k^\mu \p_\mu {\cal F}_{(0)}(x,k) = 0.
\eel
This is nothing else but the kinetic equation to be satisfied by the scalar coeffficient of the Wigner function. Equation \rfn{eq:iP1} confirms that the axial-vector coefficient is orthogonal to $k$ in both the zeroth and first orders. Doing straightforward algebraic manipulations we can check that \rf{eq:iV1} is satisfied provided \rf{eq:kineqF0} holds.

Equation~\rfn{eq:iA1} leads directly to the kinetic equation obeyed by the axial-vector coefficient
\bel{eq:kineqA0}
k^\mu \p_\mu \, {\cal A}^\nu_{(0)} (x,k) = 0, 
\quad k_\nu \,{\cal A}^\nu_{(0)} (x,k) = 0.
\eel
Using \rf{eq:kineqA0} and the orthogonality condition \rfn{eq:iP1} we can check now that \rf{eq:iS1} is also satisfied.

\subsection{Second order} \label{sec:second}

By studying the zeroth and first orders of the semiclassical expansion we have found that the basic coefficient functions are the scalar and axial-vector components. Their leading-order terms ${\cal F}_{(0)}(x,k)$ and ${\cal A}^\nu_{(0)} (x,k)$ satisfy the kinetic equations \rfn{eq:kineqF0} and \rfn{eq:kineqA0}. The axial vector coefficient should be (in the zeroth and first orders) orthogonal to the four-vector~$k$. If the functions ${\cal F}_{(0)}(x,k)$ and ${\cal A}^\nu_{(0)} (x,k)$ are known, all other coefficient functions in the zeroth order can be determined through the algebraic relations \rfn{eq:rP0}, \rfn{eq:rV0a} and \rfn{eq:rS0a}. 

We emphasize that although the system of equations derived above is consistent up to the first order in $\hbar$ (the property demonstrated in several previous studies), it is not sufficient to determine the first-order coefficient functions. We are missing dynamic equations that could be used to determine the evolution of the coefficient functions ${\cal F}_{(1)}(x,k)$ and ${\cal A}^\nu_{(1)} (x,k)$. This is expected, since we have just seen that the zeroth order is not sufficient to determine the evolution of the functions ${\cal F}_{(0)}(x,k)$ and ${\cal A}^\nu_{(0)} (x,k)$ --- this requires going to the first order. Thus, the functions ${\cal F}_{(1)}(x,k)$ and ${\cal A}^\nu_{(1)} (x,k)$ should be obtained from the analysis of the second order. Such an analysis is completely analogous to that done in the first order and, in fact, leads to the same form of equations:
\bel{eq:kineqF1}
k^\mu \p_\mu {\cal F}_{(1)}(x,k) = 0,
\eel
\bel{eq:kineqA1}
k^\mu \p_\mu {\cal A}^\nu_{(1)} (x,k) = 0, 
\quad k_\nu {\cal A}^\nu_{(1)} (x,k) = 0.
\eel
If ${\cal F}_{(1)}$ and ${\cal A}^\nu_{(1)}$ are determined, 
the quantities ${\cal P}^{(1)}$, ${\cal V}^{(1)}_\mu$, and ${\cal S}^{(1)}_{\mu \nu}$ are obtained from Eqs.~\rfn{eq:rP1a}, \rfn{eq:rV1a}, and \rfn{eq:rS1a}, respectively.

\section{Exact solutions} \label{sec:exact}

It is very interesting to observe that the algebraic structure of the equilibrium coefficient functions, defined by Eqs.~\rfn{eq:Wid1}--\rfn{eq:Wid7}, is consistent with the zeroth-order equations obtained from the semiclassical expansion of the Wigner function discussed in Sec.~\ref{sec:zeroth}, see Eqs.~\rfn{eq:rF0}--\rfn{eq:iS0}. This suggests that the global and extended global equilibrium distributions can be indeed constructed from the functions \rfn{eq:Weqpxk2} and \rfn{eq:Weqmxk2}, provided they fulfill in addition the kinetic equations \rfn{eq:kineqF0} and \rfn{eq:kineqA0}.

We have to emphasize here, however, that the equilibrium coefficient functions defined by Eqs.~\rfn{eq:FEeqpm}--\rfn{eq:SEeqpm} specify only the leading order terms in $\hbar$ of the ``true'' equilibrium function that solves the kinetic equation~\footnote{ Our approach is based on the form postulated in Ref.~\cite{Becattini:2013fla} that may be missing some important quantum contributions. In particular, the functions $\Weqxk$ are always on the mass shell, hence, they neglect off-shell quantum propagation of particles.}. To summarize our findings we can write:
\beq
{\cal F}^{(0)} &=& {\cal F}_{\rm eq},
\label{eq:FC1} \\
{\cal P}^{(0)} &=&0,
\label{eq:PC1} \\
{\cal V}^{(0)}_\mu  &=& {\cal V}_{\rm eq, \mu},
\label{eq:VC1}\\
{\cal A}^{(0)}_\mu  &=& {\cal A}_{\rm eq, \mu},
\label{eq:AC1}\\
{\cal S}^{(0)}_{\mu \nu} &=& {\cal S}_{{\rm eq}, \mu \nu},
\label{eq:SC1} \\ \bigskip
\eeq
in the zeroth order, and similarly:
\beq
{\cal P}^{(1)}  &=& -\frac{1}{2m} \, \p^\mu  {\cal A}_{\rm eq, \mu},
\label{eq:rP1EQ} \\
{\cal V}^{(1)}_\mu &=& \frac{1}{m} \left(k_\mu {\cal F}^{(1)} 
- \frac{1}{2} \p^\nu {\cal S}_{\rm eq, \nu \mu} \right),
\label{eq:rV1EQ} \\
{\cal S}_{\mu \nu}^{(1)} &=& \frac{1}{2m} \left(\p_\mu {\cal V}_{\rm eq, \nu} - \p_\nu {\cal V}_{\rm eq, \mu} \right) 
- \frac{1}{m} \epsilon_{\mu \nu \alpha \beta} k^\alpha {\cal A}_{(1)}^\beta,
\label{eq:rS1EQ}
\eeq
in the first order. 

Let us check now the constraints imposed on the equilibrium coefficient functions by Eqs.~\rfn{eq:kineqF0} and \rfn{eq:kineqA0}. One can easily find that they lead to the equations:
\bel{eq:kineqFC1}
k^\mu \p_\mu {\cal F}_{\rm eq}(x,k) = 0,
\eel
\bel{eq:kineqAC1}
k^\mu \p_\mu \, {\cal A}^\nu_{\rm eq} (x,k) = 0, 
\quad k_\nu \,{\cal A}^\nu_{\rm eq}(x,k) = 0.
\eel
Using Eqs.~\rfn{eq:FEeqpm} and \rfn{eq:AEeqpm} in  Eqs.~\rfn{eq:kineqFC1} and \rfn{eq:kineqAC1}   we conclude that the kinetic equations are exactly fulfilled if the $\beta_\mu$ field is the Killing vector defined by Eqs.~\rfn{eq:Killing} and ~\rfn{eq:Killingsol}, while the parameter $\xi$ and the spin polarization tensor $\omega_{\mu\nu}$ are constant (this implies that the parameter $\zeta$ defined by \rf{eq:zeta} is also constant). 

Consequently, the kinetic equations considered in this work (and also in the previous works that used the same mathematical setup) do not constrain the spin polarization tensor $\omega_{\mu\nu}$ to be equal to the thermal vorticity $\varpi_{\mu\nu}$. In the semiclassical approach discussed here, both tensors should be constant but may be not related to each other. This situation corresponds to extended global equilibrium rather than to global equilibrium. Most likely, the equality of the tensors $\omega_{\mu\nu}$ and $\varpi_{\mu\nu}$ (the fact expected on very general thermodynamic grounds, see Sec.~\ref{sec:global}) could follow from the proper entropy maximization. The present approach, however, does not offer any reliable method for such a calculation. We note that the first-order equations~\rfn{eq:kineqF1} and \rfn{eq:kineqA1} are decoupled in our equilibrium scheme, thus, we assume below that  ${\cal F}^{(1)}(x,k)={\cal A}^{(1)}_\mu(x,k)=0$.

It is also possible that the relation $\omega_{\mu\nu} = \varpi_{\mu\nu}$ can be necessary for the collision term to vanish. The form of the latter is, however, not known. As we have mentioned above, in this work we assume that any Wigner function of the form \rfn{eq:wig_expansion}, with the coefficient functions given by Eqs.~\rfn{eq:FEeqpm}--\rfn{eq:SEeqpm}, yields a vanishing collision integral.

\section{Local conservation laws} \label{sec:con}

Having explored consequences of the assumption that the equilibrium Wigner function satisfies exactly the kinetic equation  \rfn{eq:eqforW}, we turn now to a discussion of approximate solutions. Usually, they are obtained by demanding that only certain moments of the kinetic equation \rfn{eq:eqforW} yield zero. The selection of such moments for particles with spin is, however, not obvious and one of the aims of this work is to give some insight into this problem. To set up the stage,  we discuss in this section local conservation laws, which suggest which moments of \rfn{eq:eqforW} may be relevant for construction of the hydrodynamic framework.

\subsection{Charge current} \label{sec:cc}

Expressing the charge current ${\cal N}^\alpha (x) $ in terms of the Wigner function $\Wxk$ we obtain~\cite{deGroot:1980} 
\beq
{\cal N}^\alpha (x) 
&=&  \tr \int d^4k \, \gamma^\alpha \, \Wxk
=   \int d^4k \, {\cal V}^\alpha (x,k).
\label{eq:Nalphacal1}
\eeq
In the equilibrium case we use Eqs.~\rfn{eq:VC1} and \rfn{eq:rV1EQ} for ${\cal V}^\alpha(x,k)$. In this way we find
\beq
{\cal N}^\alpha_{\rm eq} (x) &=&  N^\alpha_{\rm eq}(x) + \delta  N^\alpha_{\rm eq}(x) ,
\label{eq:Nalphacal2}
\eeq
where
\beq
N^\alpha_{\rm eq} (x)  &=&   \frac{1}{m} \int d^4k \,  k^\alpha  {\cal F}_{\rm eq} (x,k) 
 \label{eq:Nalpha}
 \eeq
 and
 \beq
\delta  N^\alpha_{\rm eq}(x) &=& - \frac{\hbar}{2m} \int d^4k \, \p_\lambda {\cal S}_{\rm eq}^{\lambda \alpha}(x,k) .
\label{eq:dNalpha}
\eeq
We have assumed here that ${\cal F}^{(1)}(x,k)=0$, which is a trivial solution of the kinetic equation~\rfn{eq:kineqF1}. 

The charge current should be conserved, which is expressed by the equation
\bel{eq:Ncon}
  \p_\alpha N^\alpha_{\rm eq}(x)  = 0.
\eel
Here we used the property $ \p_\alpha \, \delta N^\alpha_{\rm eq}(x)  = 0$, which follows from the antisymmetry of the tensor ${\cal S}_{\rm eq}^{\lambda \alpha}(x,k)$. One can check that \rf{eq:Ncon} holds in (extended) global equilibrium, due to~\rf{eq:kineqFC1}. In the (extended) local equilibrium \rf{eq:Ncon} becomes a condition for the hydrodynamic fields: $\beta_\mu(x)$, $\xi(x)$, and $\omega_{\mu \nu}(x)$ that may vary in space and time. Substituting \rf{eq:FEeqpm} into \rf{eq:Nalpha} we obtain
\bel{eq:Nalpha1}
N^\alpha_{\rm eq} = 4\cosh(\zeta) \sinh(\xi)  
\int \frac{d^3p}{(2\pi)^3 E_p} \, p^\alpha
\,e^{-\beta \cdot p },
\eel
which agrees with Eq.~(12) from \rfc{Florkowski:2017ruc}. Doing the integral over the momentum, one finds that the charge current is proportional to the flow vector,
\bel{Nmu}
N^\alpha_{\rm eq} = n u^\alpha,
\eel
where 
\bel{nden}
n = 4 \, \cosh(\zeta) \sinh(\xi)\, \n0(T)
\eel
is the charge density~\footnote{One should include also the contribution from \rf{eq:dNalpha} to the charge current. We intend to analyze this issue in a separate paper~\cite{Florkowski:2018}. }. Here $\n0(T) = \langle(u\cdot p)\rangle_0$ is 
the number density of spin-0, neutral Boltzmann particles, obtained using the thermal average
\bel{avdef}
\langle \cdots \rangle_0 \equiv \int \f{d^3p}{(2\pi)^3 E_p}  (\cdots) \,  e^{- \beta \cdot p}.
\eel

\subsection{Energy-momentum and spin  tensors} \label{sec:tmunu}

\subsubsection{GLW formulation} \label{sec:slmnGLW}
Adopting the kinetic-theory framework derived by de Groot, van Leeuwen, and van Weert in Ref.~\cite{deGroot:1980}, where the energy-momentum tensor is expressed directly by the trace of the Wigner function,  we can use the following expression
\bel{eq:tmunu1}
T^{\mu\nu}_{\rm GLW}(x)=\frac{1}{m}\tr \int d^4k \, k^{\mu }\,k^{\nu }\Wxk=\frac{1}{m} \int d^4k \, k^{\mu }\,k^{\nu } {\cal F}(x,k).
\eel
In the equilibrium case, we consider \rf{eq:tmunu1} up to the first order in $\hbar$ using Eq.~\rfn{eq:FC1} and setting ${\cal F}^{(1)}(x,k)=0$, similarly as in the case of the charge current. Hence, with the help  of \rf{eq:FEeqpm} we obtain
\bel{eq:tmunu2}
T^{\mu\nu}_{\rm GLW}(x)=4 \cosh (\zeta) \cosh (\xi) \int \frac{d^3p}{(2 \pi )^3 E_p}p^{\mu }p^{\nu }e^{-\beta \cdot p}.
\eel
In this way we reproduce the perfect-fluid formula given earlier in \rfc{Florkowski:2017ruc},
\bel{Tmn}
T^{\mu\nu}_{\rm GLW}(x) &=& (\varepsilon + P ) u^\mu u^\nu - P g^{\mu\nu},
\eel
where the energy density and pressure are given by the expressions
\bel{enden}
\varepsilon = 4 \, \cosh(\zeta) \cosh(\xi) \, \e0(T)
\eel
and
\bel{prs}
P = 4 \, \cosh(\zeta) \cosh(\xi) \, \P0(T),
\eel
respectively. In analogy to the density $\n0(T)$, we define the auxiliary 
quantities $\e0(T) = \langle(u\cdot p)^2\rangle_0$ and $\P0(T) = -(1/3) \langle \left[ p\cdot p - (u\cdot p)^2 \right] \rangle_0$.  The energy-momentum tensor should be conserved, hence we demand
\bel{eq:Tcon}
\p_\alpha T^{\alpha\beta}_{\rm GLW}(x) = 0.
\eel
Similarly to the case of the charge conservation, one can check that \rf{eq:Tcon} holds in (extended) global equilibrium, provided~\rf{eq:kineqFC1} is satisfied. Again, in the (extended) local equilibrium \rf{eq:Tcon} becomes a condition (strictly speaking, four equations) for the hydrodynamic fields: $\beta_\mu(x)$, $\xi(x)$, and $\omega_{\mu \nu}(x)$ .

The GLW spin tensor has the following form~\cite{deGroot:1980}
\bel{eq:Smunulambda_de_Groot1}
S^{\lambda , \mu \nu }_{\rm GLW} =\frac{\hbar}{4} \, \int d^4k \, \tr \left[ \left( \left\{\sigma ^{\mu \nu },\gamma ^{\lambda }\right\}+\frac{2 i}{m}\left(\gamma ^{[\mu }k^{\nu ]}\gamma ^{\lambda }-\gamma ^{\lambda }\gamma ^{[\mu }k^{\nu ]}\right) \right) \Wxk \right].
\eel
For dimensional reasons, we have implemented here the Planck constant. Its presence implies that in equilibrium we may take the leading order expression for the Wigner function and assume $ \Wxk=\Weqxk$. Using \rftwo{eq:Weqpxk2}{eq:Weqmxk2} in \rf{eq:Smunulambda_de_Groot1}, performing the appropriate traces, and then carrying out the integration over $k$ we get
\beq
S^{\lambda , \mu \nu }_{\rm GLW}&=&\frac{\hbar \sinh (\zeta) {\cosh}(\xi)}{m^2\zeta }\int dP \, e^{-\beta \cdot p} p^{\lambda } \left(m^2\omega ^{\mu\nu}+2 p^{\alpha }p^{[\mu }\omega ^{\nu ]}{}_{\alpha } 
\right)  \label{eq:Smunulambda_de_Groot22} \nn  \\
&=& \frac{\hbar w}{4 \zeta} u^\lambda \omega^{\mu\nu} + \frac{2 \hbar \sinh (\zeta) {\cosh}(\xi)}{m^2\zeta} s^{\lambda , \mu \nu }_{\rm GLW} ,
\label{eq:Smunulambda_de_Groot2}
\eeq
where we have introduced the spin density $w$ defined by the expression~\cite{Florkowski:2017ruc}
\bel{eq:w}
w = 4 \sinh(\zeta) \cosh(\xi) n_{(0)}(T),
\eel
the auxiliary tensor
\beq
s^{\lambda , \mu \nu }_{\rm GLW} = Au^{\lambda}u^{\alpha}u^{[\mu }\omega ^{\nu ]}{}_{\alpha }+ B\left(\Delta ^{\lambda \alpha }u^{[\mu }\omega ^{\nu ]}{}_{\alpha }+u^{\lambda }\Delta ^{\alpha [\mu }\omega ^{\nu ]}{}_{\alpha }+u^{\alpha }\Delta ^{\lambda [\mu }\omega ^{\nu ]}{}_{\alpha}\right),
\eeq
and the thermodynamic coefficients
\beq 
B=-\frac{1}{\beta} \left(\varepsilon_{(0)}+P_{(0)}\right), ~~~A=\frac{1}{\beta}\left[3 \varepsilon_{(0)}+\left(3 + \frac{m^2}{T^2}\right) P_{(0)}\right]=-3B+\frac{m^2}{T}P_{(0)}.
\eeq
Since, the energy-momentum tensor derived in Ref.~\cite{deGroot:1980} is symmetric, the spin tensor \rfn{eq:Smunulambda_de_Groot2} should be also conserved (see, for example,  \rf{eq:spin_ang_mntm})
\beq
\p_\lambda S^{\lambda , \mu \nu }_{\rm GLW}(x) = 0 .
\label{eq:SGLWcon}
\eeq
This formula implies that the angular-momentum conservation holds separately for the orbital and spin parts. 

At this point, it is interesting to stress that  the coefficient function ${\cal F}_{\rm eq}(x,k)$ involves all hydrodynamic variables, i.e., $\beta_\mu$, $\xi$, and the spin polarization tensor $\omega_{\mu \nu}$ --- altogether 11 independent functions. This makes the system of Eqs.~\rfn{eq:Ncon} and \rfn{eq:Tcon} insufficient to determine their space-time dependence unless some other information is taken into account. One possibility is to assume local equilibrium state as defined in the end of Sec.~\ref{sec:bconcepts} (the third point). In this case the spin polarization tensor is equal to the thermal vorticity and the number of independent equations becomes equal to the number of unknown functions. However, since the spin polarization tensor depends on the space-time gradients of the field $\beta_\mu$ in this case, the conservation laws become second-order partial differential equations. Clearly, they do not resemble standard hydrodynamic equations and it is not obvious at the moment how one can treat and solve them. Another possibility is to introduce extended local equilibrium (the fourth point discussed in the end of Sec.~\ref{sec:bconcepts}) and to treat  the spin polarization tensor and thermal vorticity as independent quantities. The evolution of the $\omega_{\mu\nu}$ components should follow from the angular momentum conservation, which for the case discussed in this section is  reduced to \rf{eq:SGLWcon}. This approach has been proposed originally in \rfc{Florkowski:2017ruc} with a phenomenological version of the spin tensor that agrees with the first term in the second line of \rf{eq:Smunulambda_de_Groot2}.

\subsubsection{Canonical version} \label{sec:slmnCAN}

The canonical forms of the energy-momentum and spin tensors, $T^{\mu\nu}_{\rm can}(x)$ and $S^{\lambda , \mu \nu }_{\rm can}(x)$, can be obtained directly from the Dirac Lagrangian by applying the Noether theorem~\cite{Itzykson:1980rh}:
\bel{eq:tmunu1can1}
T^{\mu\nu}_{\rm can}(x)= \int d^4k \,k^{\nu } {\cal V}^\mu(x,k)
\eel
and
\beq
S^{\lambda , \mu \nu }_{\rm can}(x) &=& \frac{\hbar}{4} \, \int d^4k \,\text{tr}\left[ \left\{\sigma ^{\mu \nu },\gamma ^{\lambda }\right\} \Wxk  \right] \nn \\
&=& \frac{\hbar}{2} \epsilon^{\kappa \lambda \mu \nu} \int d^4k \, {\cal A}_{ \kappa}(x,k) \equiv \frac{\hbar}{2} \epsilon^{\kappa \lambda \mu \nu} \, {\cal A}_{ \kappa}(x).
\label{eq:Smunulambda_canonical1}
\eeq
Here we have used the anticommutation relation $\left\{\sigma ^{\mu \nu },\gamma ^{\lambda }\right\} = -2 \epsilon^{\mu\nu\lambda\kappa} \gamma_\kappa \gamma_5$ to express directly the canonical spin tensor by the axial-vector coefficient function ${\cal A}_{\kappa}(x,k)$. 

 Including the components of ${\cal V}^\mu(x,k)$ up to the first order in the equilibrium case we obtain
\bel{eq:tmunu1can2}
T^{\mu\nu}_{\rm can}(x) = T^{\mu\nu}_{\rm GLW}(x) + \delta T^{\mu\nu}_{\rm can}(x) 
\eel
where
\bel{deltaTmunu}
\delta T^{\mu\nu}_{\rm can}(x)  = -\frac{\hbar}{2m} \int d^4k k^\nu \partial_\lambda {\cal S}^{\lambda \mu}_{\rm eq}(x,k) = -\partial_\lambda S^{\nu , \lambda \mu }_{\rm GLW}(x).
\eel
The canonical energy-momentum tensor should be exactly conserved, hence, in analogy to \rf{eq:Tcon} we require 
\bel{eq:Tconcan}
\p_\alpha T^{\alpha\beta}_{\rm can}(x) = 0.
\eel
It is interesting to observe that the conservation laws \rfn{eq:Tcon} and \rfn{eq:Tconcan} are consistent, since $\partial_\mu \, \delta T^{\mu\nu}_{\rm can}(x) = 0$. The latter property follows directly from the definition of $\delta T^{\mu\nu}_{\rm can}(x) $, see \rf{deltaTmunu}.

For the equilibrium spin tensor it is enough to consider the axial-vector component in \rf{eq:Smunulambda_canonical1}  in the zeroth order, $ {\cal A}^{(0)}_{ \kappa}(x,k)= {\cal A}_{\rm eq,  \kappa}(x,k)$. Then, using \rf{eq:AEeqpm} in \rf{eq:Smunulambda_canonical1} and carrying out the integration over the four-momentum $k$ we get
\beq
S^{\lambda , \mu \nu }_{\rm can}  &=& \frac{\hbar \sinh(\zeta)\cosh(\xi)}{\zeta} \int dP \, e^{-\beta \cdot p}\left(\omega ^{\mu \nu } p^{\lambda}+\omega ^{\nu \lambda } p^{\mu}+\omega ^{\lambda \mu } p^{\nu}\right) \nn \\
&=& \frac{\hbar w}{4 \zeta} \left( u^\lambda \omega^{\mu\nu} + 
u^\mu \omega^{\nu \lambda} + u^\nu \omega^{\lambda \mu}
\right) \nn \\
&=& S^{\lambda , \mu \nu }_{\rm GLW} + S^{\mu , \nu \lambda }_{\rm GLW}+ S^{\nu , \lambda \mu }_{\rm GLW},
\label{eq:Smunulambda_canonical2}
\eeq

It is interesting to notice that the energy-momentum tensor \rfn{eq:tmunu1can2} is not symmetric.
In such a case, the spin tensor is not conserved and its divergence is equal to the difference of the energy-momentum components. For the case discussed in this section we obtain
\beq
\p_\lambda S^{\lambda , \mu \nu }_{\rm can}(x) = T^{\nu\mu}_{\rm can} - T^{\mu\nu}_{\rm can}
= -\partial_\lambda S^{\mu , \lambda \nu }_{\rm GLW}(x) + \partial_\lambda S^{\nu , \lambda \mu }_{\rm GLW}(x). 
\label{eq:Scancon}
\eeq
One can immediately check, using the last line of \rf{eq:Smunulambda_canonical2}, that \rf{eq:Scancon} is consistent with the conservation of the spin tensor in the GLW approach.

\subsubsection{Pseudo-gauge transformation} \label{sec:PsG}


In the last section we have discussed the energy-momentum and spin tensors obtained from the canonical formalism and related them to the expressions introduced by de Groot, van~Leeuven, and van~Weert.  In this section we demonstrate that the two versions of tensors are connected by a pseudo-gauge transformation. Indeed, if we introduce the tensor $\Phi^{\lambda, \mu\nu}$ defined by the relation
\bel{Phi}
\Phi^{\lambda, \mu\nu} \equiv S^{\mu , \lambda \nu }_{\rm GLW}-S^{\nu , \lambda \mu }_{\rm GLW},
\eel
we can write
\bel{psg1}
S^{\lambda , \mu \nu }_{\rm can}= S^{\lambda , \mu \nu }_{\rm GLW} -\Phi^{\lambda, \mu\nu}
\eel
and
\bel{psg2}
T^{\mu\nu}_{\rm can} = T^{\mu\nu}_{\rm GLW} + \frac{1}{2} \left(
\Phi^{\lambda, \mu\nu}+\Phi^{\mu, \nu \lambda} + \Phi^{\nu, \mu \lambda} \right).
\eel
Here, we have used the property that both $S^{\lambda , \mu \nu }_{\rm GLW} $ and $\Phi^{\lambda, \mu\nu} $ are antisymmetric with respect to exchange of the last two indices. Equations \rfn{psg1} and \rfn{psg2} are an example of the pseudo-gauge transformation discussed widely in the literature~\cite{Hehl:1976vr}. The most common use of such a transformation is connected with a change from the canonical formalism to the Belinfante one \cite{Belinfante:1940} --- it provides a symmetric energy-momentum tensor and eliminates completely the spin tensor. In a very recent work, it has been argued that the use of tensors that differ by the pseudo-gauge transformation leads to different predictions  for measurable quantities such as spectrum and polarization of particles~\cite{Becattini:2018duy}. The results presented in this work can be useful to study  such effects in more detail within explicitly defined hydrodynamic models.

\subsubsection{Hydrodynamics from moments of the kinetic equations} \label{sec:hydro}

In this section we analyze finally the issue connected with the construction of the hydrodynamic framework from the kinetic theory, namely, we try to answer the question which moments of the kinetic equations should be included to derive hydrodynamic equations. As far as we concentrate on the charge, energy, and momentum conservations, the answer is known --- we should consider the zeroth and first moments of the kinetic equation
\bel{eq:kineqFt}
k^\mu \p_\mu {\cal F}_{\rm eq}(x,k) = 0.
\eel
In this way we obtain \rf{eq:Ncon} and \rf{eq:Tcon}.

In any case, the conservation laws for charge, energy, and momentum are not sufficient to determine the dynamics of spin and they should be supplemented by information coming from the equation for the axial coefficient of the equilibrium Wigner function. The latter can be rewritten in the following form
\beq
0 &=& k^\alpha  \p_\alpha  \, 
\,\int dP\,e^{-\beta \cdot p }\, \frac{\sinh(\zeta)\, }{\zeta} 
\left[ \delta^{(4)}(k-p) e^{\xi}
+ \delta^{(4)}(k+p) e^{-\xi} \right]
\, \tilde{\omega }_{\mu \nu}\,p^{\nu} .
\label{eq:h0} 
\eeq
If we multiply the first line of \rf{eq:h0} by the four-vector $k_\beta$, contract it with the Levi-Civita tensor $\epsilon^{\mu\beta\gamma\delta}$, and then integrate the resulting equation again over $k$, we obtain the conservation of the spin tensor in the GLW version, see \rf{eq:SGLWcon}.~\footnote{We recall that in the derivation of the hydrodynamic equations we do not assume that the kinetic equations are fulfilled but expect that their specific moments vanish. We also note that the choice of the moments is not obvious. Some hints in this respect can be obtained, for example, by comparing exact solutions of the kinetic equations with the hydrodynamic equations, for example, see Ref.~\cite{Tinti:2015xra}. } This observation suggests that the form of the spin tensor derived by de Groot, van Leeuwen, and van Weert is, in fact, a very natural choice for the hydrodynamic treatment of spin. This would also indicate that one should make an attempt to derive hydrodynamic equations with spin using the GLW expression for the spin tensor. This can be done in the similar way as in Ref.~\cite{Florkowski:2017ruc}. However, it is not obvious at the moment how \rf{eq:Smunulambda_de_Groot2} can be included in a consistent construction of the hydrodynamic picture~\cite{Florkowski:2018}.

We close this section with a remark concerning the hydrodynamic equations used in \cite{Gao:2012ix}. Equations (13) and (14) from this work imply that the flow vector $u^\mu$ satisfies the Killing equation, hence it is constant (see the end of Appendix \ref{sec:Killing}). Consequently, the vorticity considered in this work is zero and no conclusions about the vorticity-polarization coupling can be drawn from the analysis presented in \cite{Gao:2012ix}.

\section{Summary and conclusions} \label{sec:summary}

In this work we have compared thermodynamic and kinetic approaches used to study relations between the spin polarization tensor and fluid vorticity in systems consisting of spin-$\onehalf$ particles. We have first discussed the thermodynamic approach that refers to general properties of global thermal equilibrium with a rigid-like rotation. Such a framework demonstrates directly that the spin-polarization and thermal-vorticity tensors are indeed equal in global equilibrium (for asymmetric energy-momentum tensors). Then, we have turned to the discussion of the kinetic approach based on the concept of the semiclassical expansion of the Wigner function. We have analyzed in more detail the case where the Wigner functions satisfy kinetic equations with a vanishing collision term.  We have found, in contrast to many earlier claims found in the literature, that this approach does not imply a direct relation between the thermal vorticity and spin polarization, except for the fact that the two should be constant in global equilibrium (we have dubbed this state an extended global equilibrium). 

Finally, we have outlined procedures for obtaining hydrodynamic equations from the kinetic equations with spin.  In the GLW case the energy-momentum tensor is symmetric and the spin tensor is conserved, while in the canonical case the energy-momentum tensor has an antisymmetric part and the spin tensor is not conserved. Nevertheless, in these two cases the total angular momentum is always conserved. We have also found that the two approaches are connected by the pseudo-gauge transformation, which we have explicitly constructed. This observation opens up new perspectives for studies of hydrodynamics with spin. From a broader point of view we notice that the classical part of the canonical energy-momentum tensor is symmetric, hence, it is suitable for the use in the context of general theory of relativity, which is a classical theory. 

Our results fill the gap between two apparently different approaches to study polarization. They indicate the importance of inclusion of the collision term in the kinetic calculations involving the Wigner function. This may shed light on the form of the equilibrium distribution (Wigner) functions in connection with the entropy production processes. The open question remains to what extent the equilibrium distributions functions used in this work remain a good approximation to more accurate, quantum equilibrium Wigner functions (with particles being not necessarily on the mass shell).

\acknowledgments

We thank F. Becattini and E. Speranza for many illuminating discussions. This work was supported in part by the Polish National Science Center Grant No. 2016/23/B/ST2/00717.

\appendix
\section{Killing equation} \label{sec:Killing}
In this section, for completeness of our presentation, we demonstrate that all solutions of ~\rf{eq:Killing} have the form \rfn{eq:Killingsol} with $b_\mu$ and $\varpi_{\mu \nu}$ being constant. We start by differentiation of~\rf{eq:Killing} with respect to coordinate $x^\alpha$. In this way we obtain
\beq
\beta_{\nu, \mu \alpha} + \beta_{\mu, \nu \alpha} = 0,
\label{eq:Killing1}
\eeq
where $, \mu \alpha$ denotes differentiation first with respect to the coordinate $x^\mu$ and then with respect to the coordinate $x^\alpha$. Changing $\nu \to \alpha$, $\mu \to \nu$, and $\alpha \to \mu$, we rewrite \rf{eq:Killing1} as
\beq
\beta_{\alpha, \nu \mu} + \beta_{\nu, \alpha \mu} = 0.
\label{eq:Killing2}
\eeq
Changing $\alpha \to \mu$, $\nu \to \alpha$, and $\mu \to \nu$ in \rf{eq:Killing2} we find
\beq
\beta_{\mu, \alpha \nu} + \beta_{\alpha, \mu \nu} = 0.
\label{eq:Killing3}
\eeq
Introducing the notation: $a = \beta_{\nu, \mu \alpha}$, $b = \beta_{\mu, \nu \alpha}$, and $c = \beta_{\alpha, \mu\nu}$, and using the fact that mixed derivatives are equal, Eqs.~\rfn{eq:Killing1}--\rfn{eq:Killing3} can be rewritten as a simple system of algebraic equations: $a+b=0, c+a=0$, and $b+c=0$, which has the solution $a=b=c=0$. This implies that the field $\beta_\mu$ is a linear function of the coordinates $x^\nu$, 
\beq
\beta_\mu = b_\mu + \varpi_{\mu \nu} x^\nu.
\label{eq:Killingsol1}
\eeq
Using the Killing equation \rfn{eq:Killing} we finally find that $\varpi_{\mu \nu}$ is antisymmetric, $\varpi_{\mu \nu} = - \varpi_{\nu \mu}$.

The $\beta_\mu$ field is usually defined by the ratio $u_\mu/T$, where $T$ is a local temperature. Thus, in the case $T$=const.  \rf{eq:Killing} implies that the four-velocity $u_\mu$ itself should be a Killing vector. Writing the solution of this equation as $u_\mu = u_\mu^0 + \alpha^0_{\mu \rho} x^\rho$, where $\alpha^0_{\mu \rho} = -\alpha^0_{\rho \mu}$ is an antisymmetric tensor with constant components, and using the normalization condition for the four-velocity, we find that: $u_\mu^0 u^\mu_0 =1$, $u^0_\mu \alpha_0^{\mu\rho}=0$, and $\alpha^0_{\rho \mu} \alpha_0^{\mu \tau}=0$. These equations imply that $\alpha^0_{\mu \nu} =0 $, which can be easily checked first in the frame where $u^0_\mu = (1,0,0,0)$. If the tensor $\alpha^0_{\mu\nu}$ vanishes in this frame it means that it is zero in all other frames. Consequently, the flow is not vortical in this case.

\section{Traces of gamma matrices} \label{sec:trgammas}

In this section we collect useful results on the traces of  products of the Dirac matrices which appear in our formalism. We use the Itzykson-Zuber conventions with $\tr \left(\gamma_5 \gamma^\alpha \gamma^\beta \gamma^\gamma \gamma^\delta \right) = -4 i \epsilon^{\alpha \beta \gamma \delta}$, where $\epsilon^{0123} = +1$~\cite{Itzykson:1980rh}. The identities used to obtain the spinor decomposition of the equilibrium Wigner functions are:
\bel{eq:idf}
\tr \left[(\slashed{p}\pm m) \SmunuU (\slashed{p} \pm m) \right] =0,
\eel
\bel{eq:idp}
\tr \left[\gamma_5 (\slashed{p} \pm m) \SmunuU (\slashed{p} \pm m) \right] =0,
\eel
\bel{eq:idv}
\tr \left[ \gamma^\alpha (\slashed{p} \pm m) \SmunuU (\slashed{p} \pm m) \right] =0,
\eel
\bel{eq:ida}
\tr \left[\gamma_\alpha  \gamma_5 (\slashed{p} \pm m) \Sigma_{\rho\sigma} (\slashed{p} \pm m) \right] = \pm 4 m \,p^\beta \, \epsilon_{\beta \alpha \rho \sigma},
\eel
and
\bel {eq:ids}
\tr \left[ 2 \Sigma^{\alpha \beta} (\slashed{p} \pm m) \Sigma^{\mu\nu}(\slashed{p} \pm m) \right]&=&4 m^2\left(g^{\alpha \mu } g^{\beta \nu }-g^{\alpha \nu } g^{\beta \mu }\right) \nn\\
&& + \, 4 \left(g^{\alpha \nu} p^\beta p^\mu 
- g^{\alpha \mu} p^\beta p^\nu + p^\alpha p^\nu g^{\beta \mu} 
- p^\alpha p^\mu g^{\beta \nu} \right).
\eel
To derive \rf{eq:ids} it is useful to use
\bel {eq:idsaux1}
\tr \left(\gamma^\alpha \gamma^\beta \,\slashed{p}\, \gamma^\mu \gamma^\nu \,\slashed{p} \right) &=& 
2 m^2 \left(g^{\alpha \beta} g^{\mu \nu} 
- g^{\alpha \mu} g^{\beta \nu} 
+ g^{\alpha \nu} g^{\beta \mu}  \right) \nn \\
&& + \,8\, \left( g^{\alpha \mu} p^\beta p^\nu 
- g^{\alpha \nu} p^\beta p^\mu 
+ p^\alpha p^\mu g^{\beta \nu} 
- p^\alpha p^\nu g^{\beta \mu} \right)
\eel
and
\bel {eq:idsaux2}
\tr\left[\Sigma^{\alpha \beta}\Sigma^{\mu\nu}\right]
= g^{\alpha \mu } g^{\beta \nu }-g^{\alpha \nu } g^{\beta \mu }.
\eel



%
%
%
%
%
%
%
%
\bibliography{pv_ref}{}
\bibliographystyle{JHEP}
\end{document}